\documentclass[aps,reprint,amsmath, amssymb,groupedaddress,floatfix]{revtex4-1}
\usepackage{natbib}
\usepackage{graphicx}
\usepackage{bm}
\usepackage{color,soul}
\usepackage{hyperref}
\usepackage{float}
\usepackage{csquotes}
\usepackage{lipsum, babel}

\usepackage{color,soul}
\usepackage{hyperref}
\hypersetup{colorlinks=true, urlcolor=blue, citecolor=blue}
\usepackage{longtable}

\newcommand{\BZT}{{\rm Ba(Ti$_{1-x}$,Zr$_x$)O$_3$}}
\newcommand{\BTO}{{\rm BaTiO$_3$}}
\usepackage{relsize}
\DeclareRobustCommand{\element}[1]{\@element#1\@nil}
\def\@element#1#2\@nil{%
  #1%
  \if\relax#2\relax\else\MakeLowercase{#2}\fi}
\begin{document}

\title{Machine learning reveals memory of the parent phases in ferroelectric relaxors \BZT}
\author{Adriana Ladera$^1$}
\email{adrianalader@usf.edu}
\author{Ravi Kashikar$^2$}
\email{ravik@usf.edu}
\author{S. Lisenkov$^2$}
\author{I. Ponomareva$^2$}
\email{iponomar@usf.edu}
\affiliation{1. Department of Computer Science and Engineering, University of South Florida, Tampa, Florida 33620, USA} 
\affiliation{2. Department of Physics, University of South Florida, Tampa, Florida 33620, USA}

\begin{abstract}

Machine learning has been establishing its potential in multiple areas of condensed matter physics and materials science. Here we develop and use an unsupervised machine learning workflow within a framework of first-principles-based atomistic simulations 
to investigate phases, phase transitions, and their structural origins in   ferroelectric relaxors, \BZT.  
We first demonstrate the applicability of the workflow to identify phases and phase transitions in the parent compound, a prototypical ferroelectric \BTO. We then apply the workflow on \BZT\, with $x\leq0.25$ to reveal (i) that some of the compounds bear a subtle memory of \BTO\, phases  beyond the point of the pinched phase transition, which could contribute to their enhanced electromechanical response;  (ii) the existence of peculiar phases with delocalized precursors of nanodomains-- likely candidates for the controversial polar nanoregions; and (iii) nanodomain phases for the largest concentrations of $x$.

\end{abstract}
  \maketitle

\section*{Introduction}


Ferroelectrics are materials that exhibit spontaneous polarization, which is reversible by the application of an external electric field \cite{10.1093/acprof:oso/9780198507789.001.0001}. These materials crystallize in polar space groups. They are also pyroelectric and piezoelectric and often feature large pyroelectric, and piezoelectric coefficients along with large dielectric constants. These responses peak in the vicinity of phase transitions. As a consequence, for practical applications it is desirable to work close to transition temperatures. 
Ferroelectric relaxors, or relaxors, differentiate themselves from normal ferroelectrics in a variety of ways. Firstly, most of them do not undergo the phase transitions associated with the change in the space group. Yet, at high temperatures relaxors exhibit paraelectric behavior, similar to the one of normal ferroelectrics. It is believed that at the Burns temperature, $T_B$, dynamic polar nanoregions (PNRs) are formed in a relaxor. As temperature decreases, their dynamics slow down and they are assumed to freeze at the  temperature $T_f$. Below this temperature the phase is nonergodic, but in many cases it can be converted into the ferroelectric one through application of an electric field. Subsequent heating results in the transition into an ergodic relaxor phase at the temperature T$_C$, which is close to $T_f$  \cite{Ahn2016, Rvw_1, https://doi.org/10.1111/j.1551-2916.2011.04952.x, doi:10.1142/9789811210433_0003}. 

It is further believed that these peculiar dynamic PNRs give origin to the hallmark features of relaxors: (i) a large and very broad peak in the temperature dependence of the dielectric constant, which is not  associated with a phase transition (unlike the case of normal ferroelectrics); (ii) and frequency dependence of the temperature associated with the maximum of the dielectric constant.  Although the PNR model is very successful in describing these and other unique features of relaxors, their  direct observation either in experiments or in atomistic simulations is challenging.  Indeed, not only are these PNRs expected to be of nanoscale size and polarized in random directions, they are also dynamic and could have frequency in the range of MHz to THz \cite{https://doi.org/10.1002/adfm.201801504, Wang2016}. 
 These are often  the time/length scales that are too short/small for experimental resolution, yet are too long/large for atomistic simulations. As a result the subject of PNRs remains under debate for many years \cite{https://doi.org/10.1002/adfm.201801504,PNG,10.1038/nmat2196}.

Relaxors are found among materials with compositional disorder, which could be both heterovalent and isovalent \cite{Ahn2016, Rvw_1}. The latter one is quite attractive  as the relaxor behavior is not overcomplicated by the  charge disorder. The \BZT\, solid solution is one such example \cite{https://doi.org/10.1111/j.1551-2916.2011.04952.x,Dixit2006,Tanmoy2011}. This family of solid solutions is attractive because the relaxor properties onset in a gradual fashion as Zr substitutes Ti  and it is also a lead free compound.  Atomistic simulations revealed that  Zr substitution of Ti strongly favors short-range repulsive force, while 
the bigger size of Zr locally favors the long-range interactions along O-Ti-O-Ti-O chains \cite{PhysRevB.97.174108}.  
 In Ref.\cite{PhysRevB.99.064111} first-principles and classical finite-temperature simulations were employed to reveal two competing effects which are primarily  responsible for the phase evolution in this family of solid solutions. They are the chemical pressure that 
Zr exerts on the \BTO\, matrix and the ferroelectric “inactivity” of Zr itself. The former
one has a stabilizing effect on ferroelectricity, while the latter one disrupts the ferroelectric cooperation. Ref. \cite{PhysRevB.102.224109} proposed that in \BZT\,  the hallmark relaxor features  persist even in the dynamically poled structures, but do not necessarily originate from PNRs dynamics.

In recent years, machine learning (ML) has emerged as a powerful research tool for condensed matter physics and materials science, which allows to extend, deepen,  and some times even revolutionize insight from both experiments and computations \cite{Schmidt2019, 10.1038/nphys4053, RevModPhys.91.045002}. 
 For example, artificial neural networks  can be trained to  identify phases and phase transitions and detect multiple types of order parameters, as well as highly non-trivial states with no conventional order, directly from raw state configurations sampled with Monte Carlo \cite{Carrasquilla2017}. 
Likewise, neural networks can be trained to identify  topological phase transitions in the Kitaev chain, the thermal phase transition in the classical Ising model, and the many-body-localization transition in a disordered quantum spin chain \cite{Carrasquilla2017}.
Another noteworthy example is the neural-network representation of density functionality theory (DFT) based  potential-energy surfaces, which provides the energy and forces as a function of all atomic positions in systems of arbitrary size and is several orders of magnitude faster than DFT \cite{PhysRevLett.98.146401}. Likewise, Gaussian approximation machine-learned potential for silicon was found to accurately reproduce DFT reference results for a wide range of observable properties, including crystal, liquid, and amorphous bulk phases, as well as point, line, and plane defects \cite{PhysRevX.8.041048}. 
These successes inspire us to find out whether ML techniques could provide an additional dimension to the ever-active investigation of microscopic origin of unusual features of relaxors.  In particular, the questions we raise are: Can ML algorithms succeed in  probing PNRs? Are they suitable to investigate phase transitions or transformations in such materials or, perhaps, even in  their parent ferroelectric compounds, like \BTO? What type of ML algorithms are suitable for such investigations? Could they reveal some subtle  microscopic details  contributing to the origin of hallmark features of relaxors? 
  
So far, ML was used to analyze the multidimensional experimental data sets describing relaxation to voltage and thermal stimuli, producing the temperature-bias phase diagram for the $(1 -x)$Pb(Mg$_1/3$Nb$_2/3$)O$_3$-$x$PbTiO$_3$ relaxor. It was concluded that ML can be used to determine phase transitions in ferroelectrics, providing a general  and robust approach toward determining the presence of critical regimes and phase boundaries \cite{Lieaap8672}. A combination of contact Kelvin probe force microscopy and ML was used to decouple the mesoscale functional response in lead lanthanum zirconate across the ferroelectric–relaxor phase transition \cite{doi:10.1021/acsami.8b15872}. Application of ML to PMN-PT solid solutions revealed a poling-like, and a relaxation-like behavior with a domain glass state \cite{griffin}.  

In this work we combine atomistic first-principles-based simulations with  unsupervised ML techniques to probe the atomistic nature of relaxor behavior as it onsets in the isovalent family of solid solutions \BZT. Our aims are (i) to demonstrate that it is possible to combine  two unsupervised ML techniques-- principal component analysis (PCA) and $K$-means clustering-- into a workflow to study or complement the study of phase transitions and transformations in ferroelectrics and ferroelectric relaxors; (ii) to reveal the subtle memory of the parent compound phases in this family of solid solutions unveiled by the ML workflow; (iii)  to present and discuss the  PNRs candidates revealed by the algorithm. 
    
\section{Computational Methodology}
To model \BZT\, we used the effective Hamiltonian proposed in Ref. \cite{PhysRevB.99.064111}. The degrees of freedom for the Hamiltonian include local soft modes which  are proportional to the local dipole moments and local strain variables that describe deformation of the unit cells. The Hamiltonian includes the interactions that are responsible for the ferroelectricity in \BZT: local mode self energy up to fourth order, harmonic short and long range interactions between the local modes, elastic deformations and interactions responsible for the electrostriction, and the term that describes interaction of local modes with the electric field.  The Hamiltonian correctly reproduces the complex phase diagram of temperature versus Zr concentration in~\BTO~\cite{PhysRevB.99.064111}.

The Hamiltonian was used in the framework of classical Molecular Dynamics (MD) simulations. We simulate Zr concentrations varying from $x =$0.05 to 0.25 in steps of 0.05. The simulations are carried out on  supercell sizes of $30\times30\times30$ of unit cells of \BTO, repeated periodically along all three Cartesian directions to simulate bulk. Each supercell contains 27,000 electric dipoles. To obtain  dipole patterns at different temperatures the simulation supercell is annealed from 450~K to 10~K in steps of 10~K. We simulated  300,000 MD steps for each temperature. The snapshots were taken every 30,000 MD steps  to produce 10 dipole patterns per temperature. A total of $P=$450  dipole patterns were harvested from the entire annealing run.  Note that each dipole pattern stores 27,000 electric dipole vectors and their positions, that is  $N=$6$\times 27,000$  values. Thus, the data can be structured into an array $X$ whose dimension is $N \times P$.
\begin{figure}
\centering
\includegraphics[width=\linewidth]{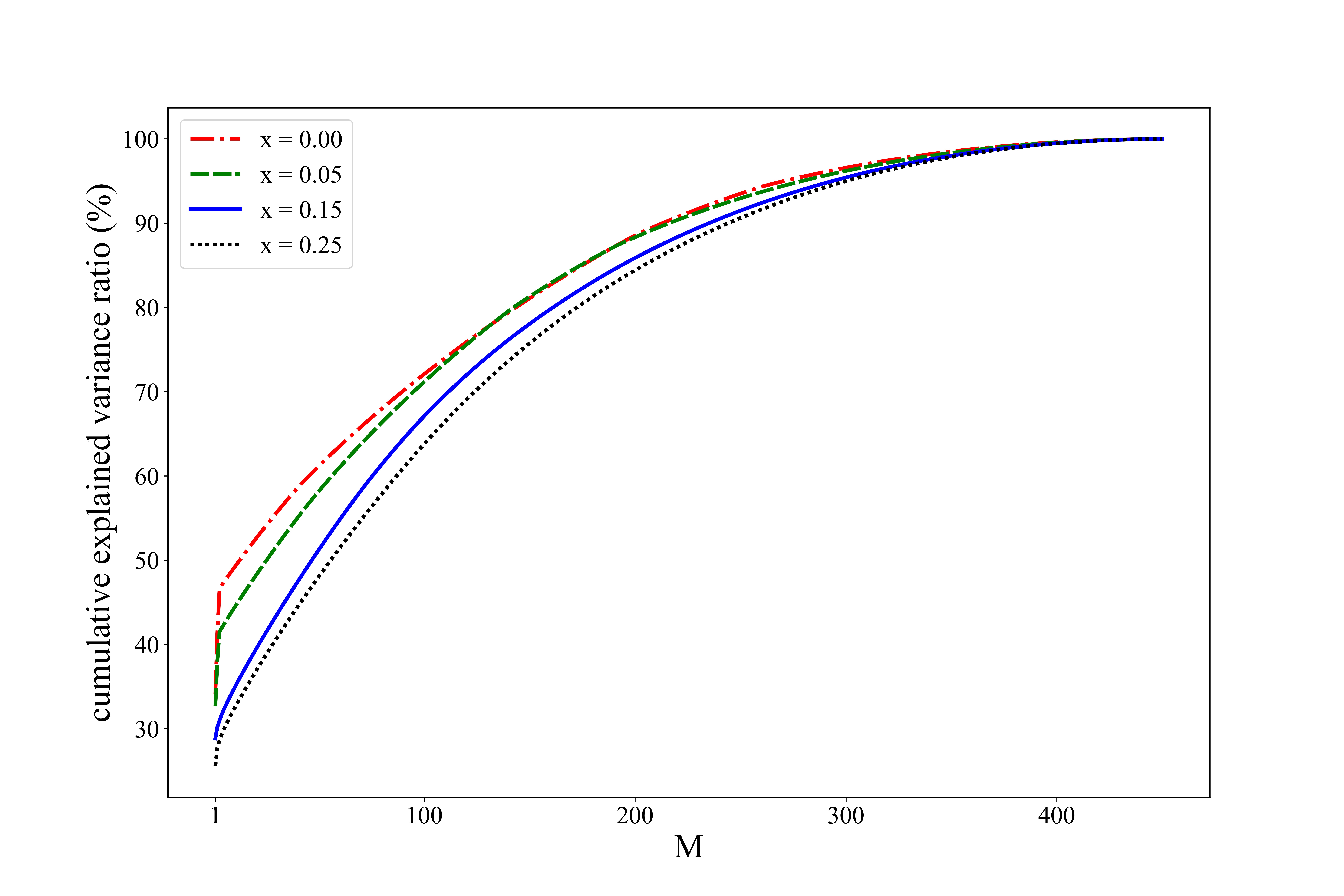}
\caption{ Cumulative explained variance ratio as a function of the number of principal components retained for select concentrations $x$.}
\label{fig_1}
\end{figure} 
Our goal is to group, or classify, the dipole patterns harvested in simulations for a given Zr concentration into different clusters using unsupervised ML. This goal will be achieved in a two-part ML workflow: (i) reduction in the dimension of the data, and (ii) the actual grouping of patterns into different clusters. For the first step we use PCA \cite{brunton_kutz_2019, watt_borhani_katsaggelos_2020}, a powerful technique for dimensionality reduction of data, especially data for unsupervised learning, as dimensionality reduction can aid in finding coherent patterns \cite{ApplicationsofPrincipalComponentAnalysistoHorticulturalResearch,8580556}. Higher dimensional data  are  projected to a lower dimensional subspace spanned by the eigenvectors of the covariance matrix, or principal components. The covariance matrix is $\frac{1}{P}X X^T$ and has a size $N\times N$. Note, that data in $X$ is the input for our PCA. They are mean-centered, that is, shifted to be symmetric with respect to the origin along each dimension.  The principal components are the eigenvectors of the covariance matrix which correspond to the largest non-negative eigenvalues. These components explain the largest percentage of variation in the data and are therefore retained, which is the essence of dimensionality reduction. 

Figure~\ref{fig_1} shows the cumulative explained variance ratio \cite{10.5555/3133359} as a function of principal components retained from simulations of different Zr concentrations.  The cumulative ratio  quantifies how much of the total variance from the original $N$-dimensional data $X$ is contained within the first $M$ principal components. We decided to keep the ratio at 99\%, which gives $M=$369, 373, 377, 380, 381, and 382 principal components for $x= $0, 0.05, 0.10, 0.15, 0.20, and 0.25 concentrations, respectively.
Thus, the size of each data point was reduced from $6\times27,000$ to  the aforementioned values of $M$.

This reduced data is then used to group $P$ patterns into $K$ clusters using $K$-means clustering \cite{MacQueen1967, 10.5555/3133359}, which is part (ii) of our ML workflow. Briefly, this ML algorithm begins with a random guess initialization of centroids (i.e. cluster centers) for $K$ clusters. Next, each data point $X_i$ for $1 \leq i \leq P$ is assigned to the nearest cluster center, that is the centroid with the shortest Euclidean squared distance from $X_i$. Given the cluster assignments, the centroids of the cluster get updated, followed by the data points reassignment between the clusters. This process repeats until the assignments no longer update within the desired tolerance or when the maximum number of iterations is reached. The procedure gives optimal data assignments to  $K$ clusters with the smallest average intra-cluster distance, that is the average distance between the points in a cluster and their centroid, over all $1 \leq j \leq K$. For all centroids, the difference between two of their consecutive iterations can be used as the tolerance for convergence and was set to $1 \times 10^{-10}$ in our case. For each  $K$ we used 10 different initial guesses for the centroid seeds with the algorithm selecting the seed that produced the best output, and set the maximum iterations value to  1600  \cite{scikit-learn}. 

Note that the $K$ value is taken as the input parameter for the algorithm. In our case we use $K$ values in the range of 1 to 10. To determine the optimal value of $K$, we utilize the elbow method \cite{10.5555/3164957}, which plots distortion and inertia as a function of $K$.  Inertia is defined as the sum of squared distances of data points to their closest centroid, whereas distortion is defined as the mean of the squared distances of the data points to their closest centroid \cite{make3020022}.
The point at which the graphs onset linear behavior is the suggested optimal $K$ value, which will be denoted as $K_O$. We implement the PCA and $K$-means clustering algorithms using the Scikit-learn software \cite{scikit-learn}. Note, that the the aforementioned values of tolerance and maximum iterations for the $K$-means clustering algorithm were empirically chosen to produce converged elbow plots.  

 \begin{figure*}
\centering
\includegraphics[width=14cm,height=15.5cm]{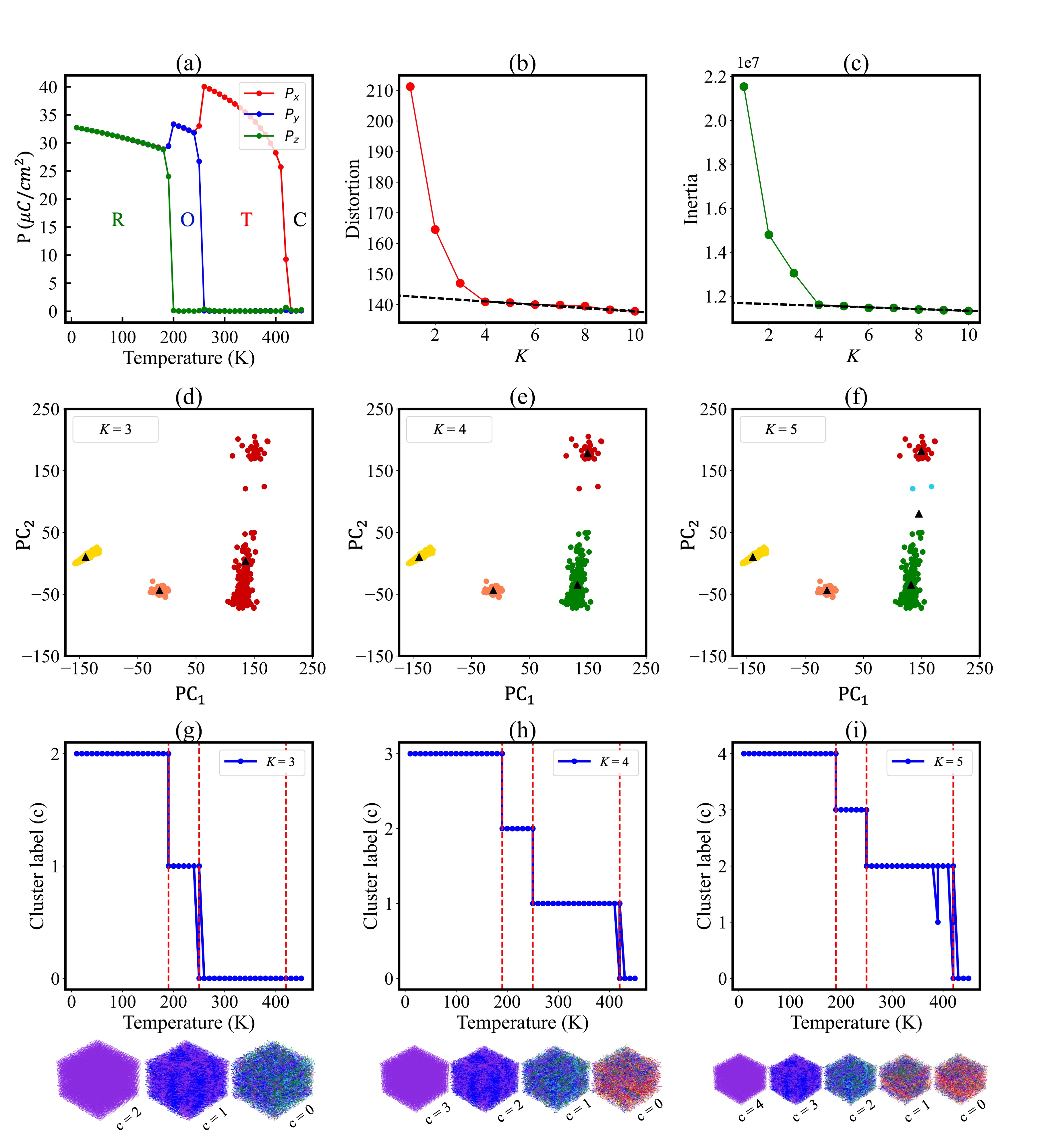}
  \caption{ Computational  data for \BTO.  Polarization components as  functions of temperature from annealing simulations (a). C, T, O, and R denote cubic, tetragonal, orthorhombic, and rhombohedral phases, respectively. Elbow plots for distortion (b) and inertia (c). The projections of dipole snapshot data onto the plane spanned by the first two principal components for different $K$ as indicated in the labels (d) - (f). The cluster-temperature relations for different $K$, as indicated in the labels (g) - (i). The insets at the bottom are the dipole patterns associated with median temperature for a given cluster. }
  \label{fig_2}
\end{figure*} 

\section{Methodology validation: the case of \element{Ba}\element{Ti}O$_3$}

We begin our investigation with \BTO, which, experimentally, exhibits three phase transitions: the paraelectric cubic to the ferroelectric tetragonal phase transition at 393~K, followed by the transition into the ferroelectric orthorhombic phase at 278~K, and finally into the ferroelectric rhombohedral phase at 183~K \cite{doi:10.1080/00150199808009173}. To identify these phase transitions in computations, we plot the components of polarization vectors (the order parameter) obtained from annealing simulations  as a function of temperature in Fig.~\ref{fig_2}(a). The different phases are clearly visible from the plot as they are associated with different polarization vectors. The transition temperatures are identified from the inflection points on the $\mathbf P(T)$ dependencies and are 420~K, 250~K and 190~K for the aforementioned phase transitions. We will refer to this way of identifying phase transitions as the thermodynamic approach.

Next we apply our ML workflow to study these phase transitions. More specifically,  the dipole patterns associated with the same phases, whose polarizations are presented in  Fig.~\ref{fig_2}(a) are first subjected to dimensionality reduction using PCA and then  grouped into different clusters using the $K$-means clustering algorithm. The distortion and inertia dependencies on $K$  used in the algorithm are given in  Fig.~\ref{fig_2}(b) and (c), respectively. Both elbow plots show the onset of linear behavior at $K=$4, which is taken to be the optimal value $K_O$. The temperature evolution of the cluster assignment  for this optimal value is given in  Fig.~\ref{fig_2}(h). Vertical lines denote the phase transition temperatures obtained from the thermodynamic approach. Note, that the clusters labels are, of course, randomly assigned in the algorithm and are re-assigned  in a way to facilitate comparison with Fig.~\ref{fig_2}(a). Figure~\ref{fig_2}(h) reveals that  steps  in the cluster-temperature relations  occur at the phase transition temperatures, which suggests that different clusters are associated  with different phases of \BTO.

\begin{figure*}
\includegraphics[width=15cm,height=15.5cm]{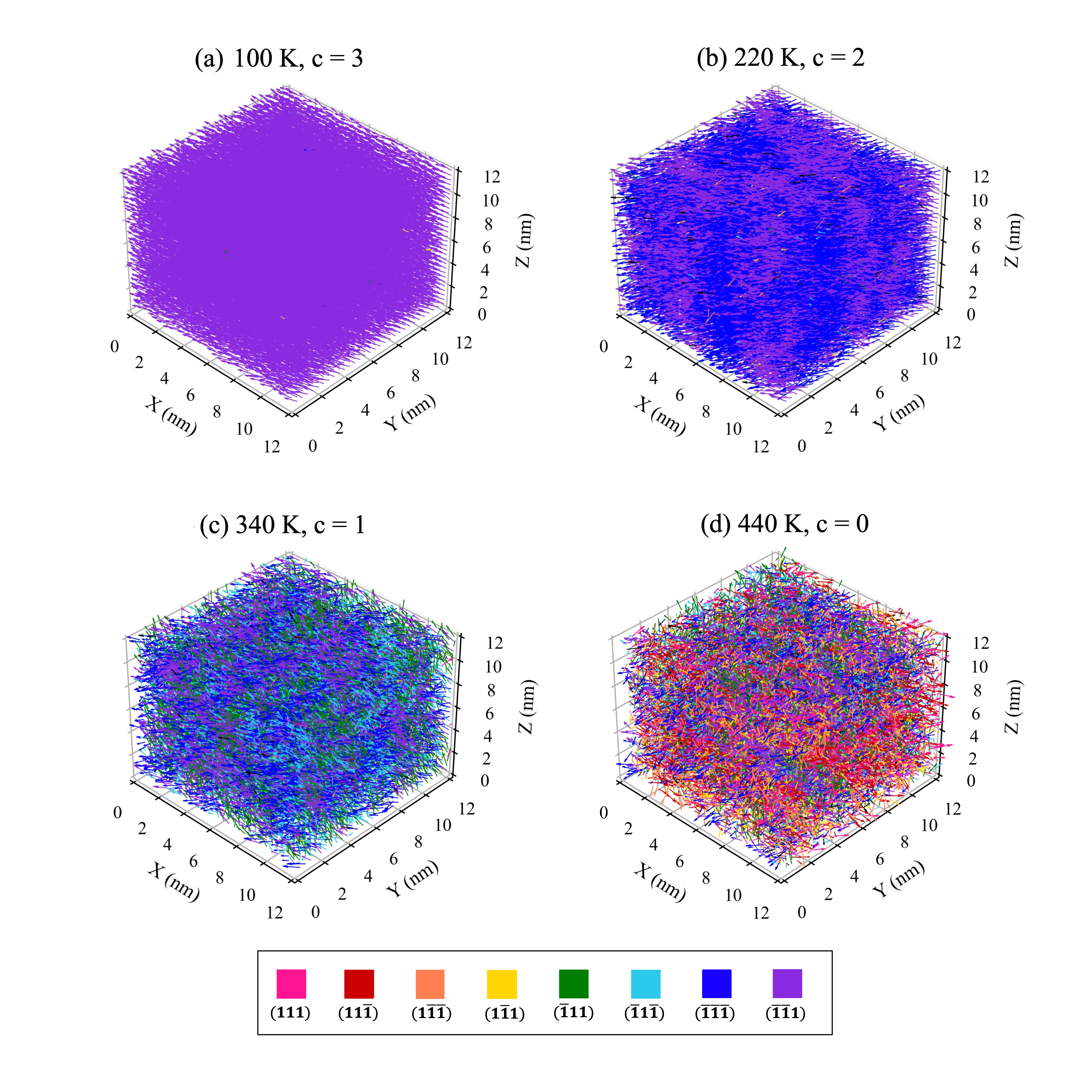}
\caption{ Snapshots of supercell dipole patterns in \BTO~at median temperatures for each  of four  clusters from $K_O=$4 computations.  }
\label{fig_3}
\end{figure*}

To further confirm that the clusters are associated with  different phases of \BTO, we inspect the dipole patterns of different clusters. Figure~\ref{fig_3} gives a representative  pattern for each cluster.  

\begin{figure*}
\centering
\includegraphics[width=17cm,height=6cm]{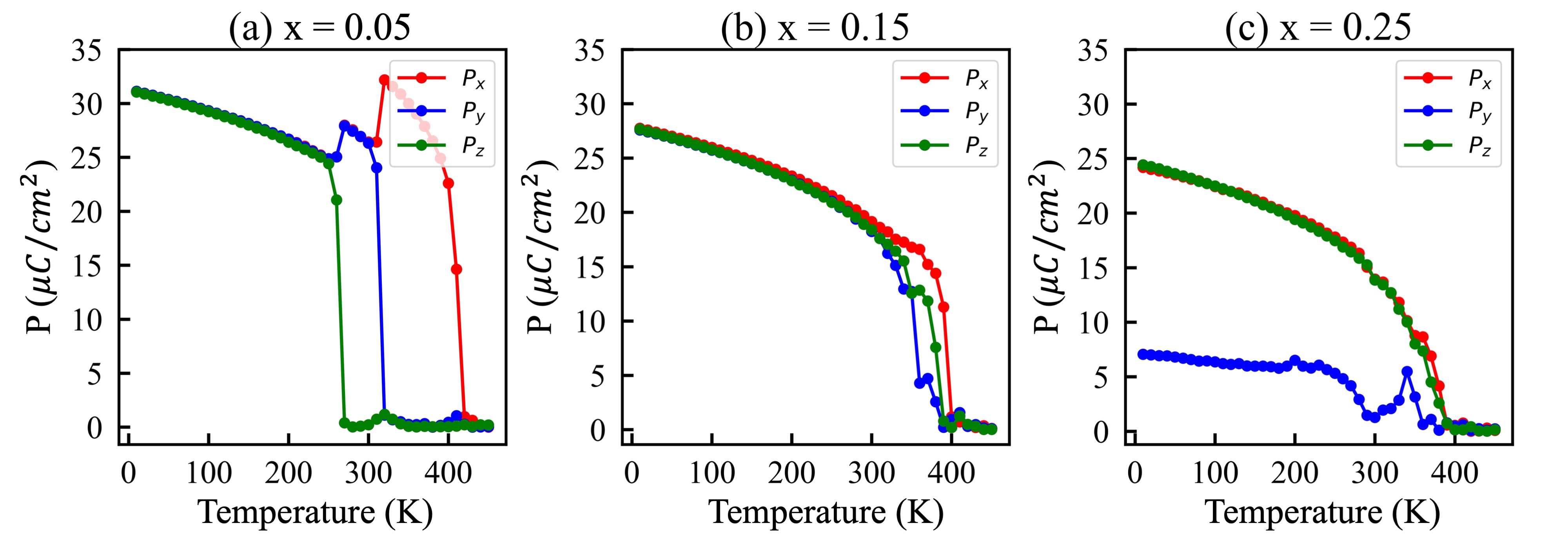}
\caption{ Components of the polarization  as a function of temperature for select concentrations $x$ in \BZT. Data for other concentrations are given in Supplementary Materials.}
\label{fig_4}
\end{figure*}

We color-code the dipoles to aid in pattern interpretation. The color-coding is inspired by the fact that the ground state of \BTO~is rhombohedral and  is associated with a [111] pseudocubic direction of the polarization along with the fact that the phase transitions in \BTO\, as well as other perovskite ferroelectrics, have both a displacive and an order-disorder component \cite{PhysRevLett.90.037601,PhysRevLett.93.037601,PhysRevB.88.064306}. Such a case is associated with shallow multiple energy wells, or grossly unharmonic single wells and is believed to be characteristic of most ferroelectrics \cite{10.1093/acprof:oso/9780198507789.001.0001}. For \BTO\, at low temperature only one out of eight [111] sites is occupied. As temperature increases the hopping between the wells onsets at the rhombohedral to orthorhombic transition to average out one of the cubic components and yield a macroscopically orthorhombic  [011] structure. Similarly, additional hopping barriers are overcome to produce a macroscopically tetragonal or cubic phase \cite{PhysRevLett.93.037601}.   There are eight [111] directions, which point along the main diagonals of the coordinate system octants. We first assign each dipole moment vector to an octant by keeping only the sign of the vector components. For example, a dipole (-$|d_x|$,$|d_y|$,-$|d_z|$) is assigned to the octant labeled ($\bar1$1$\bar1$). Each octant is assigned  its own color, as given in the legend of Fig.~\ref{fig_3}.
The dipoles with directions belonging to the same octant all have the same color, which uniquely identifies the octant. In the rhombohedral phase, for example, the dipoles point along the [111] direction and will therefore all belong to the same octant. Indeed  at the 100~K pattern (Fig.~\ref{fig_3}(a)), all dipoles have same purple color, which allows us to conclude that the patterns associated with this cluster correspond to the rhombohedral phase with the ($\bar1\bar1 1$) direction of the polarization vector. For cluster $c=$2 the median temperature pattern (Fig.~\ref{fig_3}(b))
is made up of two colors, associated with primarily ($\bar1\bar1 1$) and ($\bar1\bar1 \bar 1$) dipoles orientations, whose average results in net polarization along the ($\bar1\bar1 0$) direction, that is the orthorhombic phase. In the same manner, the median temperature  pattern for cluster $c=$1 (Fig.~\ref{fig_3}(c)) reveals the tetragonal phase with the ($\bar100$) polarization direction. Finally, the $c=$0 cluster is represented by the pattern given in Fig.~\ref{fig_3}(d) and made up of all colors, that is the fully disordered cubic phase with zero net polarization. Thus, the pattern inspection confirms that ML  clusters correspond to the different phases of \BTO, while the color-coding allows for association of patterns from different clusters with a specific color composition. This color-coding will be useful in the search for PNRs. 

It is imperative to find out how the algorithm performs outside of the optimal $K$-value. Figure~\ref{fig_2}(g) and (i) show the cluster-temperature relations for $K=$3 and 5, that is around the optimal $K_O=$4.   For $K=$3, the algorithm  merges the tetragonal and cubic phases together, while at $K=$5, an additional cluster is identified in between the tetragonal and cubic phases, likely to show the coexistence between the tetragonal and cubic phases. Indeed, the dipole distribution functions, $\rho(d_x,d_y,d_z)$, for these temperatures revealed finite densities at the  values associated with the cubic and tetragonal phases.  Thus, we conclude that $K<K_O$ does not allow to resolve phase transitions, while $K>K_O$ may improve resolution by bringing out more subtle features but can also potentially obscure identification of phase transitions. In other words, $K_O$ provides the most clear interpretation while going beyond it may bring additional insights.

To further aid in the understanding of how the algorithm classifies the patterns, we present cluster plots \cite{10.5555/3133359} in Fig.~\ref{fig_2}(d)-(f). In these plots the data are projected onto the plane spanned by the first two principal components of the mean-centered data, which are the components associated with the largest variance in the data. In our case, cumulatively, they are responsible for 42\% of the data variation. The cluster plots reveal four well-separated clusters, which are best described  with $K=$4. This further supports four as the optimal number of clusters in this case and indicates that elbow method is an appropriate approach to locating $K_O$.

\section{ML insight into \element{Ba}\element{Ti}$_{1-x}$\element{Zr}$_{x}$O$_3$}

\subsection{Application of ML workflow to \BZT}

 Having demonstrated the applicability of our approach to resolve different phases of \BTO, we next turn to \BZT. Figure~\ref{fig_4} gives the temperature evolution of polarization vector components computed for different concentrations $x$ of \BZT. We notice that the evolution of $x=$0.05 is rather similar to pure \BTO, with the main quantitative change being the transition temperatures.  The three phases merge together at $x=$0.15, which is known as the pinched phase transition \cite{https://doi.org/10.1111/j.1551-2916.2011.04952.x}. For $x=$0.25, we notice that one component of polarization is significantly smaller than the other two, suggesting the formation of nanodomains or polar nanoregions.

\begin{figure*}
\centering
\includegraphics[width=\linewidth]{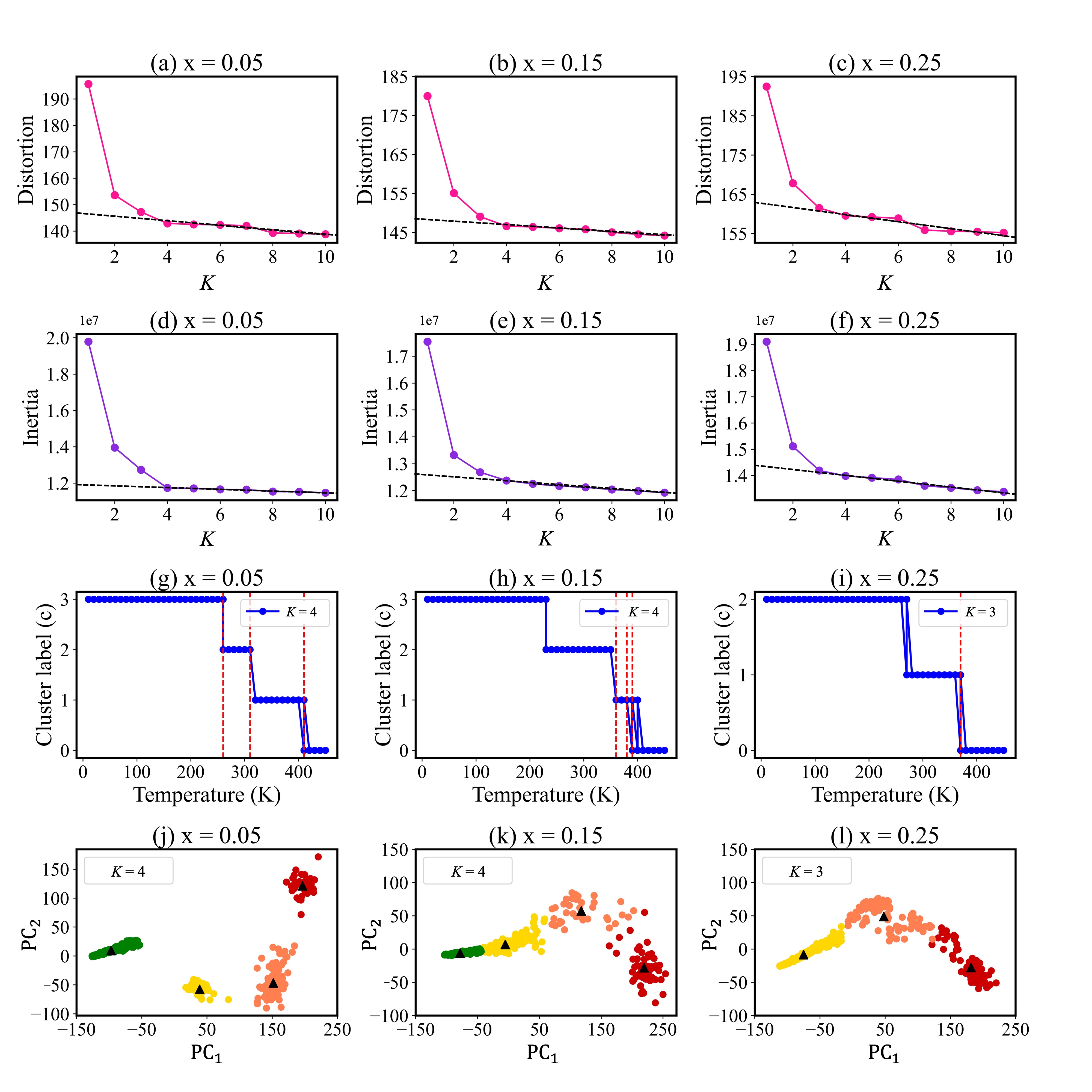}
\caption{ $K$-means clustering data for select concentrations of \BZT~as indicated in the graph titles.  Distortion elbow plots as a function of $K$ (a) - (c);  Inertia elbow plots as a function of $K$ (d) - (f);   cluster-temperature plots  for the optimal $K=K_O$ (g) - (i);   Cluster plots for $K_O$ (j) - (l).}
\label{fig_5}
\end{figure*}
 \begin{figure*}
\centering
\includegraphics[width=16cm]{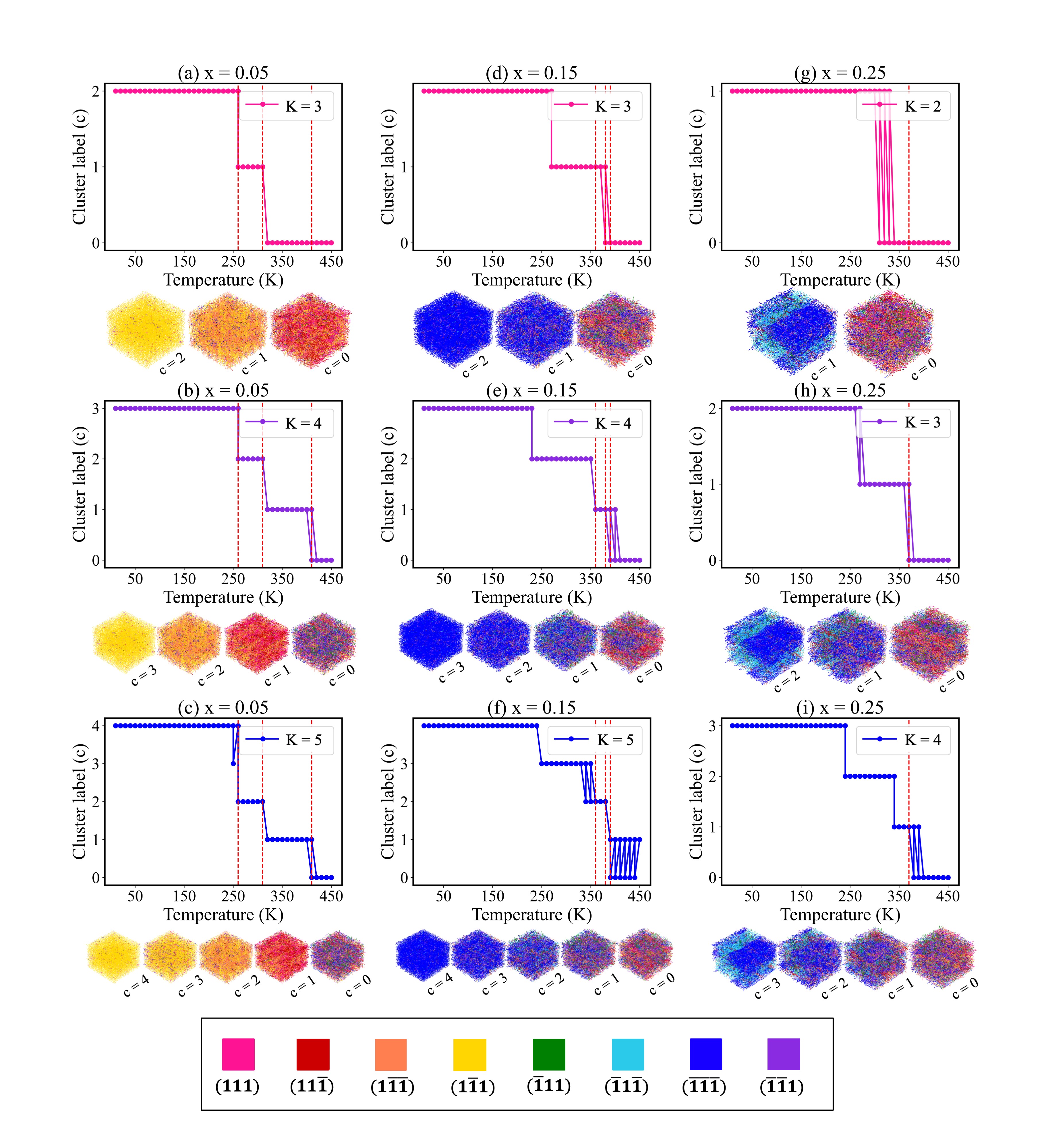}
\caption{ Cluster-temperature plots for select concentrations of \BZT\, as given in the titles and different values of $K$ given in the legends. For a given $x$ and $K$, the median temperature dipole pattern  for each cluster is given in the inset below the corresponding cluster-temperature plot. The box gives the color code assigned to the different octants of coordinate system. } 
\label{fig_6}
\end{figure*}

\begin{figure*}
\centering
\includegraphics[width=\linewidth]{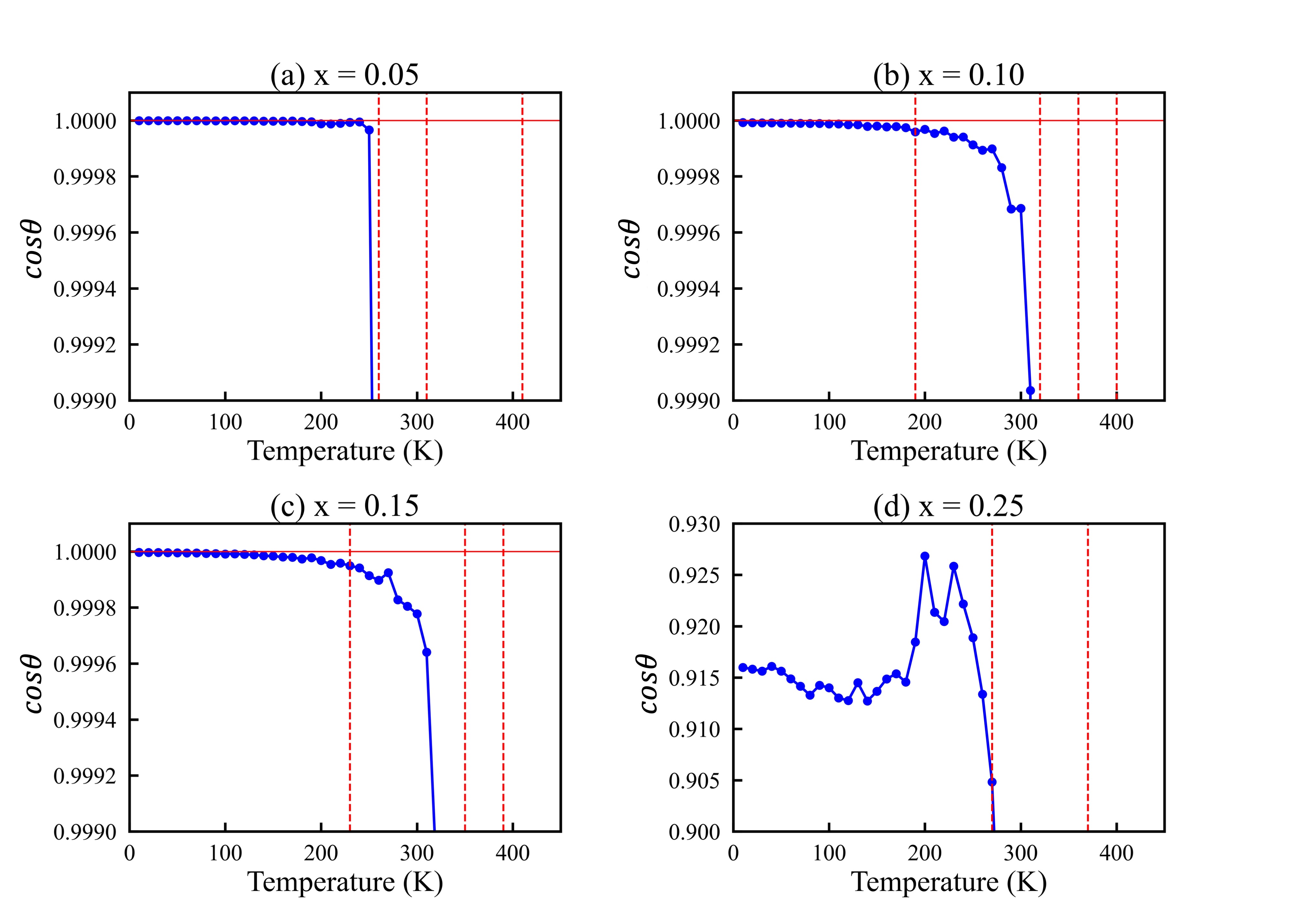}
\caption{ The cosine of the angle between polarization and the (111) direction.}
\label{fig_7}
\end{figure*}
We next apply our ML workflow  to the dipole patterns harvested in these calculations. Figure~\ref{fig_5}(a)-(f) present the elbow plots for select concentrations. We begin with $x=$0.05  where we find $K_O=$4, just as in the case of \BTO.  Quantitatively, the transition temperatures identified from the ML approach are 260~K, 320~K, and 410~K, and compare well with the values of 260~K, 310~K, and 410~K from the polarization data presented in Fig.~\ref{fig_4}(a). For $x=$0.15, the elbows in the distortion and inertia plots (Fig.~\ref{fig_5}(b) and (e)) are a bit less pronounced, yet both plots still suggest $K_O=$4. Strikingly, the cluster-temperature  relation (Fig.~\ref{fig_5}(h)) is now in dramatic contrast with the polarization evolution presented in Fig.~\ref{fig_4}(b) and does not predict the pinched phase transition. The cluster  plot  in Fig.~\ref{fig_5}(k) suggests  that the  clusters are no longer well separated.  Finally, for $x=$0.25 the elbow plots (Fig.~\ref{fig_5}(c) and (f)) are even more rounded with  the  slopes stabilizing at  $K\ge$3.  So, we take $K_O=$3 as the optimal value.  The cluster-temperature relations given in Fig.~\ref{fig_5}(i) reveal no agreement with  the predictions from the polarization data of Fig.~\ref{fig_4}(c), while the cluster plot in Fig.~\ref{fig_5}(l) suggests that the clusters are not well separated. Let us also note that for $x=$0.10 (given in Supplementary Materials), the algorithm predicts five clusters, that is $K_O$=5, whose cluster-temperature relations disagree with the polarization data in Fig.~1(a) of the Supplementary Materials. When $x=$0.20 (see Supplementary Materials) we find $K_O=$4.
Its cluster separation in the PC$_1$-PC$_2$ plane is similar to Fig.~\ref{fig_5}(k) for $x=$0.15. 

Before interpreting  these unexpected predictions from the ML algorithm we first analyze its  performance  for different $x$. In particular, we focus on the cluster assignments for  $K$ values  around the optimal $K_O$,  just as was done for the case of \BTO. Figure~\ref{fig_6} presents our data, which reveals that the increase in $K$ allows for finer separation of patterns into clusters as evident from the unique color composition  associated with each cluster.  It also suggests that the dipoles of the same color are distributed fairly uniformly in the supercell indicating a mostly homogeneous polarization field. In other words, we do not find the co-existence of different phases, domains, or PNRs in the patterns. The only exception here is $x=$0.25, where nanodomains are clearly visible for the lowest-temperature cluster. The nanodomains are associated with the regions of the same polarization direction, which are cyan and blue regions in the corresponding dipole patterns. Thus, the data on the cluster evolution with $K$ reveals that the algorithm performs well in separating patterns into different clusters and provides finer resolution as $K$ increases.  
Having established the reliability of the algorithm output we   turn to the interpretation of its unexpected predictions for the case of intermediate ($x=$0.15)  and high ($x=$0.25) Zr concentrations.

\subsection{Intermediate Zr concentrations: parent phases memory}
 Let's now investigate  the case of $x=$0.15 (Fig.~\ref{fig_5}), where the ML algorithm predicts four different clusters and therefore suggests four different phases. Recall that this prediction contradicts with the thermodynamics approach based on  $\mathbf P(T)$ evolution. Firstly, we note that the cluster-temperature relation in this case is reminiscent of that for the parent \BTO~ and suggests that this composition bears the ``memory" of the parent compound. Secondly, inspection of the color-coded patterns in Fig.~\ref{fig_6}(e) suggests that the polarization field is homogeneous, which we validated in additional analysis. That is, decomposing the snapshots given in Fig.~\ref{fig_6}(e) into single color sub-patterns, in which the dipole distribution of each color for each sub-pattern was found to be uniform.

This suggests that  the extra two clusters found in between the ones associated with cubic and rhombohedral phases do not originate from the presence of polar nanoregions or domains.  It becomes plausible that  the rhombohedral phase in the vicinity of the pinched phase transition could contain traces of the tetragonal and orthorhombic phases of the parent \BTO~and be more complex than is commonly believed. To investigate into this possibility we turn to Fig.~\ref{fig_4} and notice that in the rhombohedral phase for $x=$0.15 the three components of polarization are not exactly equal, suggestive of a lower symmetry monoclinic phase. This can be quantified by computing  the angle between the polarization in the rhombohedral phase  and the (111) direction.

Figure~\ref{fig_7} shows the deviation of the cosine of the angle from the ideal one as a function of temperature in the rhombohedral or rhombohedral-like phase  for each concentration $x$. The vertical lines separate regions associated with different clusters from the $K_O$ predicted by the ML algorithm.  Indeed, the data predict that the clusters ``inherited" from \BTO~ are associated with very subtle polarization rotation away from the ideal direction. Such rotations serve as ``memory" of the orthogonal phase of the parent compound. Polarization rotation has  been known to give origin to the enhanced electromechanical response \cite{Fu2000}. 
Therefore, the subtle polarization rotation  could contribute to the superior responses of ferroelectric relaxors.

\subsection {High Zr concentrations: nanodomains and their PNR-like precursors}

In case of $x=$0.25, the $K$-means algorithm identifies three clusters. The low temperature one exhibits nanoregions, or nanodomains, with the ($\bar1$1$\bar1$) and ($\bar1\bar1\bar1$)  polarization directions (cyan and blue regions in Fig. \ref{fig_6}(g)-(i)), while the high temperature cluster is a fully-disordered macroscopically cubic phase. The intermediate temperature cluster is associated with formation of a precursor for the nanodomains colored in blue in the $c=$2 pattern but with the surrounding matrix being disordered. This can be illustrated by the fact that red and yellow color present in the $c=$1 cluster pattern are no longer present for the $c=$2 cluster. We believe that $c=$1 cluster is a likely candidate for a PNR, although a rather delocalized one. Perhaps, such delocalization can be viewed as a percolation of multiple PNRs. It should be stressed that such percolated PNRs are ``submerged" in the ``sea" of disordered dipoles, which we found out through  color-decomposition of its pattern. We note, that we find a similar picture for the case of $x=$0.20 (presented in Supplementary Materials). The ML algorithm finds both a cluster associated with a delocalized nanodomain precursor  and the one associated with nanodomains. Nanodomains have been predicted to couple dynamically with mechanical deformations \cite{PhysRevLett.107.177601} which could also potentially be contributed to enhanced electromechanical properties.  

These findings allow us to conclude that the ML approach  is a powerful tool in resolving nontrivial and subtle  polar phases of complex ferroelectrics and relaxors with both homogeneous and inhomogeneous polarization fields.

\section{Conclusions}
In summary, we have developed and used an unsupervised ML workflow to investigate ferroelectric relaxors \BZT. We found the approach to work very well in identifying different phases of the parent compound \BTO. For \BZT, with $x=$0.10 and 0.15 the algorithm identified an additional ``rhombohedral-like"   phase overlooked by the traditional thermodynamic approach. Such a phase is associated with polarization rotating away from its ideal direction, which may  contribute to the enhanced electromechanical response of some relaxors. At $x=$0.25, the algorithm   reveals a rather unusual phase, also undetected through a traditional approach. Atomistically, the phase contains a delocalized precursor for nanodomains,  a plausible candidate for PNRs. We conclude that the ML workflow  combined with atomistic simulations is a powerful tool for investigating phases, phase transitions and their structural  origins in basic and complex ferroics.

\section{Acknowledgement}
The work is supported by the U.S. Department of Energy, Office of Basic Energy Sciences, Division of Materials Sciences and Engineering under grant DE-SC0005245.

\bibliography{paper}

\newpage
\onecolumngrid

\newpage
\begin{center}
   \textbf{\Large Supplementary Material}
\end{center}

 \vspace{1cm}
 \renewcommand{\thefigure}{S\arabic{figure}}

\setcounter{figure}{0}
 \begin{figure}[h]
\centering
\includegraphics[width=\linewidth]{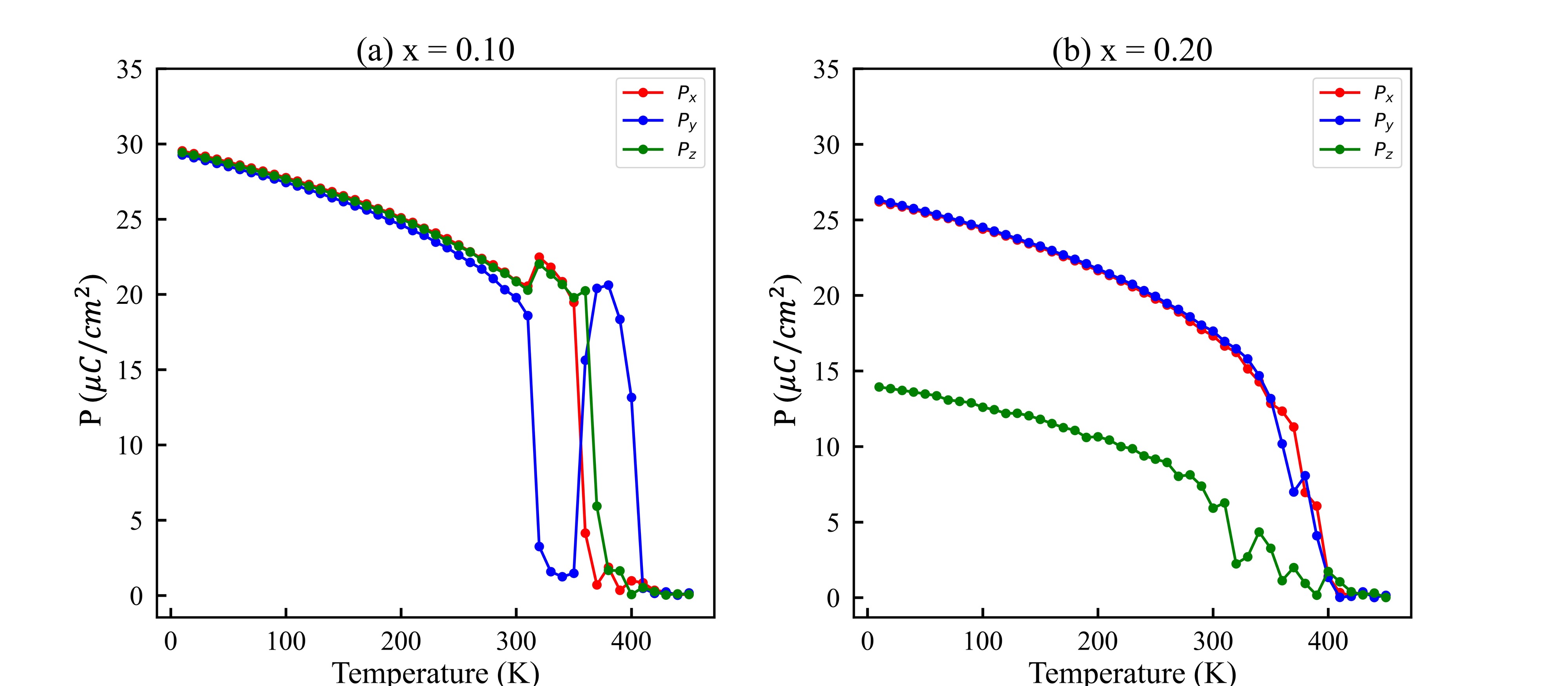}
\caption{ Polarization components as a function of temperature for $x=$0.10 and 0.20.}
\label{supp_1}
\end{figure}

 \begin{figure}
\centering
\includegraphics[width=15cm]{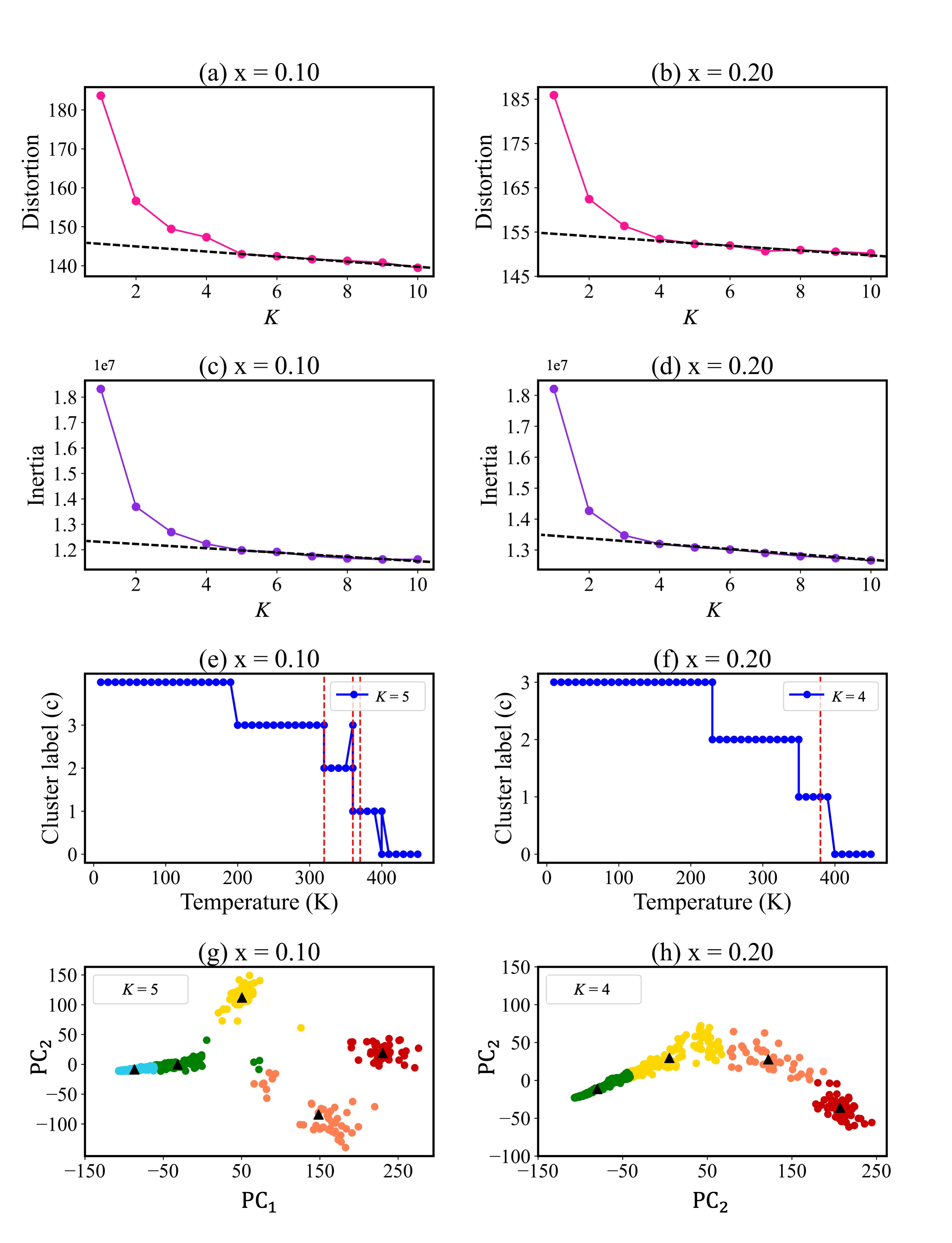}
\caption{ Data for concentrations $x=$0.10 and 0.20. (a,b) Distortion elbow plots. (c,d) Inertia elbow plots. (e,f) Cluster-temperature plots. (g,h) Cluster plots showing the relationship between $PC_1$ and $PC_2$.}
\label{supp_2}
\end{figure}

 \begin{figure}
\centering
\includegraphics[width=\linewidth]{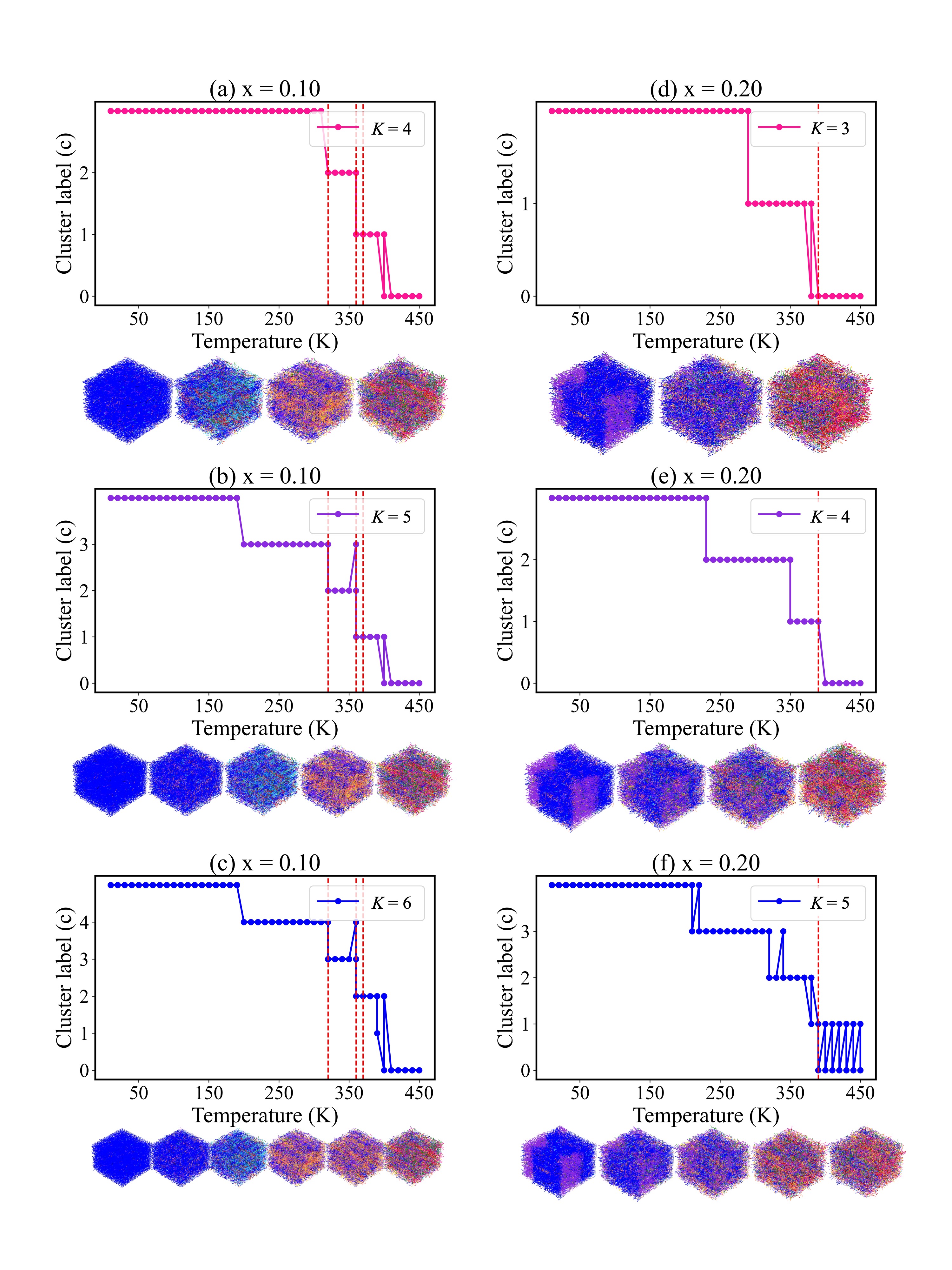}
\caption{ Cluster-temperature plots and their respective median temperature dipole patterns of each cluster.}
\label{supp_3}
\end{figure}
\end{document}